\documentclass[preprint,amsmath,amssymb,aps,eqsecnum]{revtex4-1}

\usepackage{dcolumn}
\usepackage{bm}

\begin{document}

\title{Theory of Infinitely Extended Particles}

\author{Seyyed Mahmoud Hessaby}
\altaffiliation{Editing and reproduction from original document by Sina Khorasani}
\email{khorasani@sina.sharif.edu}

\affiliation{
Department of Physics\\
University of Tehran, Tehran, Iran
}

\begin{abstract}
The difficulties with which the concept of point-like particles is beset, such as the infinities encountered in the existing theories of elementary particles, suggest a different approach to the study of these particles. Instead of restricting ourselves to the concept of point-like particles, we should extend our investigation to the implication of the concept of particles having infinite extension. Such a particle should consist of a continuous distribution of energy over all space, the energy density tending to zero at infinity.

To achieve this aim, we introduce into the theory of general relativity the postulate that the gravitational, electric and nuclear fields are special cases of a more general field. An expression is obtained for the gravitational potential which differs from the usual expression of the potential accepted in general relatvity, and which gives an energy density for the particle at every point of space, the integral of which over all space is equal to the mass of the particle, the greatest part of the mass being concentrated near the center of the spherical pattern constituting the particle. The particle is thus seen to consist of the energy of its field. No infinities are encountered in the integrations.

The same result is obtained for a charged particle. The charge density is spread out over all space and the integrals of the charge density and energy density are respectively equal to the charge and mass of the particle.

The electric potential this obtained is inserted in Dirac's wave equation, and gives a series of equations of increasing degree, the first of which gives the mass of the muon.

In addition to the expressions obtained for the electric and gravitational potentials, an expression is found for a potential which has the form of a dipole potential.  When inserted in Dirac's wave equation, this potential gives the values of the masses of baryons. When inserted in the Klein-Gordon equation, this potential gives the values of the masses of mesons.
\end{abstract}

\maketitle

\section{Foreword}

The original author of the manuscript, Professor Seyyed Mahmoud Hessaby 
\cite{1}, devised the initial theory of Continuous Particles
while he attended a visiting position at the Princeton's Institute of
Advanced Study by invitation of Albert Einstein. He published the initial
line of thoughts in 1947 in the Proceedings of National Academy of Sciences
\cite{2}. He also published a second paper in 1957 in French language
entitled Mod\`ele de particule infinie \cite{3} (Model of Infinite Particles).

About twenty years later in 1977, he managed to formulate a fully extended
and unified version of his theory but for unclear reasons it remained as
unpublished; he only reproduced it in very few numbers at Tehran University
Press \cite{4}, where he had founded himself. This TeX article has been typeset 
based on a rare remaining copy obtained from within the US market. 

This paper presents the only existing theory which unifies the three forces 
and also successfully estimates the mass ratios of various elementary particles. 
At present, there is simply no other theory being capable of estimation of 
particle mass ratios, at least within the framework of Standard Model. 

Aside from a couple of minor calculation errors and typos which have been
removed during the reproduction, Professor Hessaby's derivations seem to be 
completely flawless. 

Professor Mahmoud Hessaby's contributions to the infrastructure of
science in the Iranian society was just overwhelming. He founded the
Tehran University, Iranian Physical Society, Atomic Energy Organization
of Iran, Institute for Geophysics, the first Radio Broadcasting Center,
the first Private Hospital, and much more.

Professor Seyyed Mahmoud Hessaby passed away in 1992. He is correctly recognized 
as the \textit{Father of Modern Physics in Iran}.

\section{The Generalized Field}
\subsection{The Four-Potential}

Our postulate is that the gravitational, electric, and nuclear fields are special cases of a more general field. Designating the four potential by $\Phi_\mu$, the field in the general case is given by the expression

\begin{equation}
\label{i1}
F_{\mu\nu}=\frac{\partial \Phi_{\mu}}{\partial x_\nu}-\frac{\partial \Phi_\nu}{\partial x_\mu}
\end{equation}

\noindent
The charge-current vector is

\begin{equation}
\label{i2}
J^{\mu}=\frac{1}{2\pi}(F^{\mu\nu})_{,\nu}
\end{equation}

\noindent
and the energy tensor is

\begin{equation}
\label{i3}
U^{\nu}_{\mu}=\frac{1}{2\pi}(-F^{\nu\alpha}F_{\mu\alpha}+\frac{1}{4}g^{\nu}_{\mu}F^{\alpha\beta}F_{\alpha\beta})
\end{equation}

\noindent
By contraction of (\ref{i3}) we have also

\begin{equation}
\label{i4}
U=0
\end{equation}

We assume first the existence of only two components of $\Phi_\mu$ viz, $\Phi_3$ and $\Phi_4$, $\Phi_4$ depending only on $r$, while $\Phi_3$ may depend on both $r$ and $\theta$. The components of the field are as given in the table~\ref{tab1.1}.

\begin{table}
\caption{\label{tab1.1}Components of the field tensor $F_{\mu\nu}$.}
\begin{ruledtabular}
\begin{tabular}{cccc}
0 & 0 & $-\frac{\partial \Phi_3}{\partial x_1}$ & $-\frac{\partial \Phi_4}{\partial x_1}$ \\
0 & 0 & $-\frac{\partial \Phi_3}{\partial x_2}$ & 0 \\
$\frac{\partial \Phi_3}{\partial x_1}$ & $\frac{\partial \Phi_3}{\partial x_2}$ & 0 & 0 \\
$\frac{\partial \Phi_4}{\partial x_1}$ & 0 & 0 & 0 \\
\end{tabular}
\end{ruledtabular}
\end{table}

\subsection{The Potential $\Phi_4$}
We take first the case in which the component $\Phi_3=0$. We write the line element in the form

\begin{equation}
\label{ii1}
ds^2=-e^\alpha dr^2-e^\beta r^2 d\theta^2-e^\gamma r^2 \sin^2\theta d\phi^2+e^\delta dt^2
\end{equation}

\noindent
where we assume that $\alpha$, $\beta$, $\gamma$, and $\delta$ depend only on $r$.

In order to write down the components of $U^\nu_\mu$, we calculate first $F^{\alpha\beta}F_{\alpha\beta}$. Since we assume that $\Phi_4$ depends only on $r$, we have $\frac{\partial \Phi_4}{\partial \theta}=\frac{\partial \Phi_4}{\partial \phi}=0$, so that

\begin{equation}
\label{ii2}
F^{\alpha\beta}F_{\alpha\beta}=F^{14}F_{14}+F^{41}F_{41}=2g^{11}g^{44}(F_{14})^2=-2e^{-(\alpha+\delta)}(F_{14})^2
\end{equation}

\noindent
Also

\begin{equation}
\label{ii3}
F^{1\alpha}F_{1\alpha}=F^{4\alpha}F_{4\alpha}=g^{11}g^{44}(F_{14})^2=-e^{-(\alpha+\delta)}(F_{14})^2
\end{equation}

\noindent
so that

\begin{equation}
\label{ii4}
U^1_1=-U^2_2=-U^3_3=U^4_4=\frac{1}{4\pi}e^{-(\alpha+\delta)}(F_{14})^2
\end{equation}

\noindent
The contracted tensor is
\begin{equation}
\label{ii5}
U=\sum U^\nu_\mu=0
\end{equation}

\noindent
Designating the Ricci tensor by $R^\nu_\mu$, the expression for the energy tensor is

\begin{equation}
\label{ii6}
T^\nu_\mu=-\frac{a}{4\pi}(R^\nu_\mu-\frac{1}{2}g^\nu_\mu R)
\end{equation}

\noindent
where $a$ is a dimensional constant.

Since we have identically $U=0$, the identification of $T^\nu_\mu$ with $U^\nu_\mu$, gives by contraction

\begin{equation}
\label{ii7}
U=T=\frac{a}{4\pi}R=0
\end{equation}

\noindent
so that we have identically $R=0$.

Taking account of (\ref{ii7}), relation (\ref{ii6}) becomes

\begin{equation}
\label{ii8}
U^\nu_\mu=T^\nu_\mu=-\frac{a}{4\pi}R^\nu_\mu
\end{equation}

\noindent
relations (\ref{ii4}) and (\ref{ii8}) give

\begin{equation}
\label{ii9}
R^1_1=-R^2_2=-R^3_3=R^4_4=-\frac{1}{a}e^{-(\alpha+\delta)}(F_{14})^2
\end{equation}

\noindent
We write $R^\nu_\mu$ in terms of $\alpha$, $\beta$, $\gamma$, $\delta$. The only Christoffel symbols which do not vanish are enlisted in table~\ref{tab2.1}.

\begin{table}
\caption{\label{tab2.1}Non-vanishing Christoffel symbols.}
\begin{ruledtabular}
\begin{tabular}{ll}
$\{11,1\}=\frac{1}{2}\alpha^\prime$ & $\{21,2\}=\frac{1}{2}\beta^\prime+\frac{1}{r}$\\
$\{12,2\}=\frac{1}{2}\beta^\prime+\frac{1}{r}$ & $\{22,1\}=-e^{\beta-\alpha}(\frac{1}{2}r^{2\beta^\prime}+r)$\\
$\{13,3\}=\frac{1}{2}\gamma^\prime+\frac{1}{r}$ & $\{23,3\}=\frac{\cos\theta}{\sin\theta}$\\
$\{14,4\}=\frac{1}{2}\delta^\prime$ & \\
$\{31,3\}=\frac{1}{2}\gamma^\prime+\frac{1}{r}$ & $\{41,4\}=\frac{1}{2}\delta^\prime$\\
$\{32,3\}=\frac{\cos\theta}{\sin\theta}$ & $\{44,1\}=\frac{1}{2}\delta^\prime e^{\delta-\alpha}$\\
$\{33,1\}=-e^{\gamma-\alpha} \sin^2\theta (\frac{1}{2}r^2\gamma^\prime+r)$ & \\
$\{33,2\}=-e^{\gamma-\beta} \sin\theta\cos\theta$ & \\
\end{tabular}
\end{ruledtabular}
\end{table}

\noindent
The expressions that we obtain for the $R^\nu_\mu$ are

\begin{eqnarray}
\label{ii10}
R^1_1=e^{-\alpha} &&(-\frac{1}{2}\beta^{\prime\prime}-\frac{1}{2}\gamma^{\prime\prime}-\frac{1}{2}\delta^{\prime\prime}-\frac{1}{4}\beta^{\prime 2}-
\frac{1}{4}\gamma^{\prime 2}-\frac{1}{4}\delta^{\prime 2}+\frac{1}{4}\alpha^\prime\beta^\prime\\&&
+\frac{1}{4}\alpha^\prime\gamma^\prime+\frac{1}{4}\alpha^\prime\delta^\prime+\frac{\alpha^\prime}{r}-\frac{\beta^\prime}{r}-\frac{\delta^\prime}{r})
\nonumber
\end{eqnarray}

\begin{eqnarray}
\label{ii11}
R^2_2=e^{-\alpha}&&(-\frac{1}{2}\beta^{\prime\prime}-\frac{1}{4}\beta^{\prime 2}+\frac{1}{4}\alpha^\prime\beta^\prime-
\frac{1}{4}\beta^\prime\gamma^\prime-\frac{1}{4}\beta^\prime\delta^\prime\\&&
+\frac{1}{2}\frac{\alpha^\prime}{r}-\frac{3}{2}\frac{\beta^\prime}{r}-\frac{1}{2}\frac{\gamma^\prime}{r}-\frac{1}{2}\frac{\delta^\prime}{r}-
\frac{1}{r^2})+\frac{e^{-\beta}}{r^2}\nonumber
\end{eqnarray}

\begin{eqnarray}
\label{ii12}
R^3_3=e^{-\alpha}&&(-\frac{1}{2}\gamma^{\prime\prime}-\frac{1}{4}\gamma^{\prime 2}+\frac{1}{4}\alpha^\prime\gamma^\prime-\frac{1}{4}\beta^\prime\gamma^\prime-\frac{1}{4}\gamma^\prime\delta^\prime\\&&
+\frac{1}{2}\frac{\alpha^\prime}{r}-\frac{1}{2}\frac{\beta^\prime}{r}-\frac{3}{2}\frac{\gamma^\prime}{r}-
\frac{1}{2}\frac{\delta^\prime}{r}-\frac{1}{r^2})+\frac{e^{-\beta}}{r^2}\nonumber
\end{eqnarray}

\begin{equation}
\label{ii13}
R^4_4=e^{-\alpha}(-\frac{1}{2}\delta^{\prime\prime}-\frac{1}{4}\delta^{\prime 2}+\frac{1}{4}\alpha^\prime\delta^\prime-\frac{1}{4}\beta^\prime\delta^\prime-
\frac{1}{4}\gamma^\prime\delta^\prime-\frac{\delta^\prime}{r})
\end{equation}

By relation (\ref{ii7}) the contracted tensor $R$ is zero

\begin{eqnarray}
\label{ii14}
R=\sum R^\mu_\mu=e^{-\alpha}&&(-\beta^{\prime\prime}-\gamma^{\prime\prime}-\delta^{\prime\prime}-\frac{1}{2}\beta^{\prime 2}
-\frac{1}{2}\gamma^{\prime 2}-\frac{1}{2}\delta^{\prime 2}\\&&
+\frac{1}{2}\alpha^\prime\beta^\prime+\frac{1}{2}\alpha^\prime\gamma^\prime+\frac{1}{2}\alpha^\prime\delta^\prime
-\frac{1}{2}\beta^\prime\delta^\prime-\frac{1}{2}\gamma^\prime\delta^\prime-\frac{1}{2}\beta^\prime\delta^\prime \nonumber\\&&
+\frac{2\alpha^\prime}{r}-\frac{3\beta^\prime}{r}-\frac{3\gamma^\prime}{r}-\frac{2\delta^\prime}{r}-\frac{2}{r^2})+\frac{2e^{-\beta}}{r^2}
\nonumber\\&&=0\nonumber
\end{eqnarray}

We remark that expressions (\ref{ii11}) and (\ref{ii12}) are identical in form except for the interchange of $\beta^\prime$ and $\gamma^\prime$. This fact, together with the equality $R^2_2=R^3_3$, suggests that we assume the equality of $\beta^\prime$ and $\gamma^\prime$. Replacing $\gamma^\prime$ by $\beta^\prime$ in relations (\ref{ii10}) to (\ref{ii14}), and making use of the equality $R^1_1=R^4_4$, we obtain

\begin{equation}
\label{ii15}
e^{-\alpha}(\beta^{\prime\prime}+\frac{1}{2}\beta^{\prime 2}+2\frac{\beta^\prime}{r}-\frac{1}{2}\alpha^\prime\beta^\prime-\frac{1}{2}\beta^\prime\delta^\prime-\frac{\alpha^\prime}{r}-\frac{\delta^\prime}{r})=0
\end{equation}

\subsubsection{Case I}
We consider first the case where $\beta=\gamma=0$. Relation (\ref{ii15}) gives then $\delta^\prime=-\alpha^\prime$, and relation (\ref{ii14}), with $\delta=-\alpha$ becomes

\begin{equation}
\label{ii16}
e^\delta(-\delta^{\prime\prime}-\delta^{\prime 2}-4\frac{\delta^\prime}{r}-\frac{2}{r^2})+\frac{2}{r^2}=0
\end{equation}

\noindent
The solution

\begin{equation}
\label{ii17}
e^\delta=(1+\frac{K}{r})^2
\end{equation}

\noindent
where $K$ is a constant depending on the mass of the particle and having the dimensions of a length, satisfies equation (\ref{ii16}). In this special case, we have therefore

\begin{equation}
\label{ii18}
e^\alpha=e^{-\delta}=(1+\frac{K}{r})^{-2}
\end{equation}

\noindent
and since we have assumed $\beta=\gamma=0$, the line element becomes

\begin{equation}
\label{ii19}
ds^2=-\frac{1}{(1+\frac{K}{r})^2}dr^2-r^2d\theta^2-r^2\sin^2\theta d\phi^2+(1+\frac{K}{r})^2 dt^2
\end{equation}

\subsubsection{Case II}
If we extend the validity of the equality $\delta^\prime=-\alpha^\prime$ to the general case where $\beta^\prime\neq 0$, the differential equation (\ref{ii15}) reduces to

\begin{equation}
\label{ii20}
\beta^{\prime\prime}+\frac{1}{2}\beta^{\prime 2}+2\frac{\beta^\prime}{r}=0
\end{equation}

\noindent
which admits the solution

\begin{equation}
\label{ii21}
\beta^\prime=-\frac{2A}{r(r+A)}
\end{equation}

\noindent
giving

\begin{equation}
\label{ii22}
\beta=2\log \frac{r+A}{r}=\log(1+\frac{A}{r})^2
\end{equation}

\noindent
so that

\begin{equation}
\label{ii23}
e^\beta=(1+\frac{A}{r})^2
\end{equation}

\noindent
To find the expression for $\alpha$, we make use of the relation $R=0$ in equation (\ref{ii14}), which when set $\gamma=\beta$ becomes

\begin{equation}
\label{ii24}
e^{-\alpha}(-2\beta^{\prime\prime}-\frac{3}{2}\beta^{\prime 2}-6\frac{\beta^\prime}{r}+\alpha^{\prime\prime}-\alpha^{\prime 2}+4\frac{\alpha^\prime}{r}+2\alpha^\prime\beta^\prime-\frac{1}{r^2})+\frac{e^{-\beta}}{r^2}=0
\end{equation}

\noindent
The solution

\begin{equation}
\label{ii25}
\alpha^\prime=\beta^\prime=-\frac{2A}{r(r+A)}
\end{equation}

\noindent
satisfies relation (\ref{ii24}). We have thus, with $\delta=-\alpha$

\begin{eqnarray}
\label{ii26}
&&e^\alpha=e^\beta=e^\gamma=(1+\frac{A}{r})^2\\
&&e^\delta=(1+\frac{A}{r})^{-2}\nonumber
\end{eqnarray}

\noindent
and the line element assumes the form

\begin{equation}
\label{ii27}
ds^2=-(1+\frac{A}{r})^2(dr^2+r^2d\theta^2+r^2\sin^2\theta d\phi^2)+\frac{1}{(1+\frac{A}{r})^2}dt^2
\end{equation}

\section{The Gravitational Field}

\subsection{First Solutions of the Field Equation for $\Phi_4$}

We consider the form of the line element as found in (\ref{ii19})

\begin{equation}
\label{iii1}
ds^2=-\frac{1}{(1+\frac{K}{r})^2}dr^2-r^2d\theta^2-r^2\sin^2\theta d\phi^2+(1+\frac{K}{r})^2 dt^2
\end{equation}

\noindent
The field $F_{14}$ is given by relation (\ref{ii4}), which $\delta=-\alpha$

\begin{equation}
\label{iii2}
U^4_4=\frac{1}{4\pi}e^{-(\alpha+\delta)}(F_{14})^2=\frac{1}{4\pi}(F_{14})^2
\end{equation}

\noindent
Relation (\ref{ii13}) becomes, since we have here $\delta=-\alpha$ and $\beta=\gamma=0$

\begin{equation}
\label{iii3}
R^4_4=e^\delta(-\frac{1}{2}\delta^{\prime\prime}-\frac{1}{2}\delta^{\prime 2}-\frac{\delta^\prime}{r})
\end{equation}

\noindent
By (\ref{ii17}), we have $e^\delta=(1+\frac{K}{r})^2$, so that

\begin{eqnarray}
\nonumber
&&\delta^\prime=-\frac{2K}{r(r+K)}\\&&
\delta^{\prime\prime}=2K [\frac{1}{r^2(r+K)}+\frac{1}{r(r+K)^2} ] \nonumber
\end{eqnarray}

\noindent
and we get

\begin{equation}
\label{iii4}
R^4_4=-\frac{K}{r^4}
\end{equation}

\noindent
Relation (\ref{iii2}), together with relation (\ref{ii8}) gives then

\begin{equation}
\label{iii5}
F_{14}=\frac{\sqrt{a}K}{r^2}
\end{equation}

\noindent
The energy density (\ref{ii8}), becomes

\begin{equation}
\label{iii6}
U^{44}=-\frac{a}{4\pi}R^{44}=-\frac{a}{4\pi}g^{44}R^4_4=-\frac{a}{4\pi}e^{-\delta}R^4_4=\frac{a}{4\pi}\frac{K^2}{r^2(r+K)^2}
\end{equation}

\noindent
The integral of energy density over the whole space is

\begin{equation}
\label{iii7}
W=\frac{a}{4\pi}K^2\int{\frac{1}{r^2(r+K)^2}r^2\sin^2\theta d\theta d\phi dr}=aK^2[-\frac{1}{r+K}]^\infty_0=aK
\end{equation}

\noindent
The mass of the particle being $m$, expression (\ref{iii7}) must be equal to $mc^2$

\begin{equation}
\label{iii8}
aK=mc^2
\end{equation}

The dimension of $K$ being that of a length, the dimensions of $a$ must be that of a force. We remark that the combination of $\frac{Gm}{c^2}$, where $G$ is the gravitational constant, has the dimension of a length, and that the combination $\frac{c^4}{G}$ has the dimension of a force. We set

\begin{eqnarray}
\label{iii9}
&&a=\frac{c^4}{G}\\
&&K=\frac{Gm}{c^2}\nonumber
\end{eqnarray}

We see that in the integral of the energy density (\ref{iii7}) we encounter no infinities, and that the mass of the particle consists of the integral of its energy density over all space. The particle is seen to have infinite extension, the greatest part of its mass being nevertheless concentrated near the center of spherical pattern constituting the particle.

The line element (\ref{ii19}) becomes now

\begin{equation}
\label{iii10}
ds^2=-\frac{1}{(1+\frac{Gm}{c^2 r})^2} dr^2-r^2d\theta^2-r^2\sin^2\theta d\phi^2+(1+\frac{Gm}{c^2 r})^2 dt^2
\end{equation}

\noindent
The difference between this solution and Einstein's solution

\begin{equation}
\label{iii11}
ds^2=-\frac{1}{(1-\frac{2m}{r})}dr^2-r^2d\theta^2-r^2\sin^2\theta d\phi^2+(1-\frac{2m}{r})dt^2
\end{equation}

\noindent
is that in the line element (\ref{iii10}) the $g_{\mu\nu}$ are perfect squares, and also that they do not have any singularities apart from the origin, giving withal a finite value for the energy of the particle.

The contravariant charge current-density vector is

\begin{eqnarray}
\label{iii12}
J^4&&=\frac{1}{2\pi}(F^{\mu\nu})_{,\nu}\\&&\nonumber
=\frac{1}{2\pi\sqrt{-g}}\frac{\partial}{\partial r}(F^{4\nu}\sqrt{-g})\\&& \nonumber
=\frac{1}{2\pi r^2}\frac{\partial}{\partial r}(g^{44}g^{11}F_{41}r^2)\\&& \nonumber
=\frac{1}{2\pi r^2}\frac{\partial}{\partial r}(F_{41}r^2)\\&& \nonumber
=\frac{1}{r^2}\frac{\partial}{\partial r}(\frac{\sqrt{G}m}{r^2}r^2)\\&& \nonumber
=0
\end{eqnarray}

\noindent
The vanishing of $J^4$ is of course to be expected in the case of the gravitational field.

The interpretation of $g_{11}=-(1+\frac{Gm}{c^2 r})^{-2}$ is that of a strain in the direction of the radius vector, reducing the length of the unit mesh in that direction by the factor $(1+\frac{Gm}{c^2 r})^{-2}$. The $R_{\mu\nu}$ are not interpreted in terms of a curvature, but rather in terms of strains in space. The identity $R\equiv0$ means here that the total strain at any point of space is zero.

\section{The Electric Field and Leptons}
\subsection{Second Solution of the Field Equations for $\Phi_4$}

We now consider the line element as given in (\ref{ii27})

\begin{equation}
\label{iv1}
ds^2=-(1+\frac{A}{r})^2(dr^2+r^2d\theta^2+r^2\sin^2\theta d\phi^2)+\frac{1}{(1+\frac{A}{r})^2}dt^2
\end{equation}

The two quantities $R^\nu_\mu$ and $U^\nu_\mu$ being proportional to each other, we set as in (\ref{ii8}) and (\ref{ii4})

\begin{equation}
\label{iv2}
U^4_4=-\frac{b}{4\pi}R^4_4=\frac{1}{4\pi}e^{-(\alpha+\delta)}(F_{14})^2
\end{equation}

\noindent
By (\ref{ii13}) we have

\begin{equation}
\label{iv3}
R^4_4=e^{-\alpha}(-\frac{1}{2}\delta^{\prime\prime}-\frac{1}{4}\delta^{\prime 2}+\frac{1}{4}\alpha^\prime\delta^\prime-\frac{1}{4}\beta^\prime\delta^\prime-\frac{1}{4}\gamma^\prime\delta^\prime-\frac{\delta^\prime}{r})
\end{equation}

\noindent
Here, since we have $\alpha=\beta=\gamma=-\delta$ we get

\begin{equation}
\nonumber
R^4_4=e^\delta(-\frac{1}{2}\delta^{\prime\prime}-\frac{\delta^\prime}{r})
\end{equation}

\noindent
Replacing $e^\delta$ by $(1+\frac{A}{r})^{-2}$ we find

\begin{equation}
\label{iv4}
R^4_4=-\frac{A^2}{(r+A)^4}
\end{equation}

\noindent
so that

\begin{eqnarray}
\label{iv5}
&&R^4_4=-\frac{1}{b}(F_{14})^2=-\frac{A^2}{(r+A)^4}\\ && \nonumber
F_{14}=\sqrt{b}\frac{A}{(r+A)^2}\\ && \nonumber
U^4_4=-\frac{bA^2}{4\pi (r+A)^4}
\end{eqnarray}

The integral over all space is

\begin{equation}
\label{iv6}
\int U^4_4 dv=\frac{bA}{3}
\end{equation}

\noindent
Setting this equal to the mass of the particle, we get

\begin{equation}
\label{iv7}
\frac{bA}{3}=mc^2
\end{equation}

\noindent
This expression has the dimensions of an energy. As $A$ has the dimension of length, $b$ must have the dimensions of a force. We remark that the combination $\frac{e^2}{mc^2}$ has the dimension of a length, and the combination $\frac{m^2c^4}{e^2}$ has the dimensions of a force. We set

\begin{eqnarray}
\label{iv8}
&& b=\frac{9m^2c^4}{e^2}\\&& \nonumber
A=\frac{1}{3}\frac{e^2}{mc^2}
\end{eqnarray}

\noindent
so that

\begin{equation}
\label{iv9}
bA^2=e^2
\end{equation}

The electric field is by (\ref{iv4}) and (\ref{iv9})

\begin{equation}
\label{iv10}
E=F_{14}=\frac{\sqrt{b}A}{(r+\frac{e^2}{3mc^2})^2}=\frac{e}{(r+\frac{e^2}{3mc^2})^2}
\end{equation}

The distance $A$ is equal to

\begin{equation}
\label{iv11}
A=\frac{e^2}{3mc^2}=\frac{1}{3}\frac{e^2}{\hbar c}\frac{\hbar}{mc}=\frac{1}{3}\alpha\lambdabar=0.9393\times 10^{-13} \textrm {cm}
\end{equation}

The covariant charge current-density vector gives

\begin{equation}
\label{iv12}
J_4=\frac{1}{2\pi}(F^\nu_4)_\nu=\frac{1}{2\pi\sqrt{-g}}\frac{\partial}{\partial x_\nu}(F^\nu_4\sqrt{-g})-\frac{1}{2}\frac{\partial g^{\alpha\beta}}{\partial x_4}F_{\alpha\beta}
\end{equation}

In the static case the last term is zero, and we have, taking account of (\ref{iv1})

\begin{equation}
\label{iv13}
J_4=\frac{1}{2\pi(r+A)^2}\frac{\partial}{\partial r}[F^1_4(r+A)^2]
\end{equation}

\noindent
Since $F^1_4=g^{11}F_{41}=-(1+\frac{A}{r})^{-2}e(r+A)^2$ we get

\begin{equation}
\label{iv14}
J_4=-\frac{e}{2\pi(r+A)^2}\frac{\partial}{\partial r}\frac{1}{(1+\frac{A}{r})^2}=\frac{eAr}{\pi(r+A)^5}
\end{equation}

The integral of this expression over all space is

\begin{equation}
\label{iv15}
\int J_4 dv=4\pi\frac{eA}{\pi}[-\frac{1}{r+A}+\frac{3}{2}\frac{A}{(r+A)^3}+\frac{A^2}{(r+A)^3}+\frac{A^3}{4(r+A)^4}]_0^\infty=e
\end{equation}

\noindent
that is, the integral of the charge density over all space is equal to the charge of the particle.

We note that we do not encounter any infinities in our calculations. The mass of the particle is spread out over all space, and so is its charge. We interpret the $g_{\mu\nu}$ not in terms of a curvature, but in terms of constraints in the unit lengths at every point of space. The equality of $\alpha$, $\beta$, and $\gamma$ in the line element (\ref{iv1}), is interpreted as an isotropic constraint in all three directions at every point of space. We note that in the gravitational case the vanishing of $\gamma$ and $\beta$ in the line element (\ref{ii1}) denotes that in that case the only constraint is in the direction of the radius vector.

\subsection{The Muon}
The relativistic wave equations for a particle in a central potential are

\begin{eqnarray}
\label{v1}
&&(E+mc^2-V)F-2\pi hc\frac{dG}{dr}-\frac{2\pi hck}{r}G=0 \\ && \nonumber
(E-mc^2-V)G+2\pi hc\frac{dF}{dr}-\frac{2\pi hck}{r}F=0
\end{eqnarray}

\noindent
where $F$ and $G$ are two wave functions and $r$ is the distance to the origin. We substitute the quantities $V=-(r+\frac{e^2}{3mc^2})^{-2}e^2$, $z=\frac{E}{mc^2}$, $\beta_1=1+z$, $\beta_2=1-z$, $\beta=\sqrt{\beta_1\beta_2}=\sqrt{1-z^2}$, and the fine-structure constant $\alpha=\frac{e}{\hbar c}=0.00729735$ in equations (\ref{v1}), and introduce two new wave functions $f$ and $g$ defined by

\begin{eqnarray}
\nonumber && F(\rho)=e^{-\rho}f(\rho) \\
\nonumber && G(\rho)=e^{-\rho}g(\rho)
\end{eqnarray}

\noindent
and we obtain the following relations

\begin{eqnarray}
\label{v2}
&& g^\prime-g+\frac{kg}{\rho}-(\frac{\beta_1}{\beta}+\frac{\alpha}{\rho+\frac{\alpha\beta}{3}})f=0\\ \nonumber
&& f^\prime-f-\frac{kf}{\rho}-(\frac{\beta_2}{\beta}-\frac{\alpha}{\rho+\frac{\alpha\beta}{3}})g=0
\end{eqnarray}

We introduce two new wave functions $\phi$ and $\psi$ defined by

\begin{eqnarray}
\nonumber && f=\rho^s (\rho+\frac{\alpha\beta}{3})^t \phi \\
\nonumber && g=\rho^s (\rho+\frac{\alpha\beta}{3})^t \psi
\end{eqnarray}

\noindent
and we obtain the equations

\begin{eqnarray}
\label{v3}
\rho (\rho+\frac{\alpha\beta}{3})\psi^\prime+&&[(\rho+\frac{\alpha\beta}{3})(s+k-\rho)+t\rho]\psi\\ \nonumber &&
-[\frac{\beta_1}{\beta}(\rho+\frac{\alpha\beta}{3})+\alpha]\rho\phi=0\\ \nonumber
\rho (\rho+\frac{\alpha\beta}{3})\psi^\prime+&&[(\rho+\frac{\alpha\beta}{3})(s-k-\rho)+t\rho]\phi\\ \nonumber &&
-[\frac{\beta_2}{\beta}(\rho+\frac{\alpha\beta}{3})-\alpha]\rho\psi=0
\end{eqnarray}

\noindent
We expand $\phi$ and $\psi$ in terms of $\rho$

\begin{eqnarray}
\label{v4}
\phi=\sum a_n \rho^n \\ \nonumber
\psi=\sum b_n \rho^n
\end{eqnarray}

\noindent
and substitute these expansions in the equations (\ref{v3}); we equate to zero the constant terms

\begin{eqnarray}
\nonumber (s+k)\frac{\alpha\beta}{3}b_0=0\\
\nonumber (s-k)\frac{\alpha\beta}{3}a_0=0
\end{eqnarray}

\noindent
We must therefore have either $s=+k$ and $b_0=0$, or $s=-k$ and $a_0$=0. We choose $s=+k$ and therefore $b_0=0$.

We introduce a new variable

\begin{equation}
\nonumber
x=\rho+\frac{\alpha\beta}{3}
\end{equation}

\noindent
and substitute in the relations (\ref{v3}), and we obtain the two equations

\begin{eqnarray}
\label{v5}
x(x-\frac{\alpha\beta}{3})\psi^\prime+&&[2kx+(t-x)(x-\frac{\alpha\beta}{3})]\psi\\ \nonumber &&
-(\frac{\beta_1}{\beta}x+\alpha)(x-\frac{\alpha\beta}{3})\phi=0 \\ \nonumber
x(x-\frac{\alpha\beta}{3})\phi^\prime+&&[(t-x)(x-\frac{\alpha\beta}{3})]\phi\\ \nonumber &&
-(\frac{\beta_2}{\beta}x-\alpha)(x-\frac{\alpha\beta}{3})\psi=0
\end{eqnarray}

\noindent
We expand $\phi$ and $\psi$ in terms of $x$

\begin{eqnarray}
\nonumber \phi=\sum c_n x^n\\
\nonumber \psi=\sum d_n x^n
\end{eqnarray}

\noindent
and substitute in the equations (\ref{v5}); we equate to zero the constant terms

\begin{eqnarray}
\nonumber -t\frac{\alpha\beta}{3}d_0+\frac{\alpha^2\beta}{3}c_0=0\\
\nonumber -t\frac{\alpha\beta}{3}c_0-\frac{\alpha^2\beta}{3}d_0=0
\end{eqnarray}

\noindent
which give $t^2=-\alpha^2$, that is $t=\pm i \alpha$. We choose $t=+i\alpha$. Replacing $t$ by this value in equation (\ref{v3}), we obtain the two equations

\begin{eqnarray}
\label{v6}
\rho(\rho+\frac{\alpha\beta}{3})\psi^\prime&&+[-\rho^2+(2k+i\alpha-\frac{\alpha\beta}{3})\rho+2k \frac{\alpha\beta}{3}]\psi\\
\nonumber && -[\frac{\beta_1}{\beta}(\rho+\frac{\alpha\beta}{3})+\alpha]\rho\phi=0\\ \nonumber
(\rho+\frac{\alpha\beta}{3})\phi^\prime&&+[-\rho+i\alpha-\frac{\alpha\beta}{3}]\phi\\ \nonumber &&
-[\frac{\beta_2}{\beta}(\rho+\frac{\alpha\beta}{3})-\alpha]\psi=0
\end{eqnarray}

We substitute the expansions (\ref{v4}) into equations (\ref{v6}) and equate to zero the coefficients of successive powers of $\rho$. We obtain thus an infinite sequence of equations connecting the different coefficients $a_n$, $b_n$ of expansions (\ref{v4})

\begin{eqnarray}
\label{v7}
&&\frac{\alpha\beta}{3}a_1+(i\alpha-\frac{\alpha\beta}{3})a_0=0\\ \nonumber
&&\frac{\alpha\beta}{3}(1+2k)b_1-(\frac{\alpha\beta_1}{3}+\alpha)a_0=0\\ \nonumber
&&\frac{2\alpha\beta}{3}a_2+(1+i\alpha-\frac{\alpha\beta}{3})a_1-(\frac{\alpha\beta_2}{3}-\alpha)b_1-a_0=0\\ \nonumber
&&(2+2k)\frac{\alpha\beta}{3}b_2+(1+2k+i\alpha-\frac{\alpha\beta}{3})b_1-(\frac{\alpha\beta_1}{3}+\alpha)a_1-\frac{\beta_1}{\beta}a_0=0\\ \nonumber
&&3\frac{\alpha\beta}{3}a_3+(2+i\alpha-\frac{\alpha\beta}{3})a_2-(\frac{\alpha\beta_2}{3}-\alpha)b_2-a_1-\frac{\beta_2}{\beta}b_1=0\\ \nonumber
&&(3+2k)\frac{\alpha\beta}{3}b_3+(2+2k+i\alpha-\frac{\alpha\beta}{3})b_2-(\frac{\alpha\beta_1}{3}+\alpha)a_2-b_1-\frac{\beta_1}{\beta}a_1=0\\ \nonumber
&&\cdot\cdot\cdot\cdot\cdot\cdot\cdot\cdot\\ \nonumber
&&n\frac{\alpha\beta}{3}a_n+(n-1+i\alpha-\frac{\alpha\beta}{3})a_{n-1}-(\frac{\alpha\beta_2}{3}-\alpha)b_{n-1}-a_{n-2}-\frac{\beta_2}{\beta}b_{n-2}=0\\ \nonumber
&&(n+2k)\frac{\alpha\beta}{3}b_n+(n-1+2k+i\alpha-\frac{\alpha\beta}{3})b_{n-1}-(\frac{\alpha\beta_1}{3}+\alpha)a_{n-1}-b_{n-2}-\frac{\beta_1}{\beta}a_{n-2}=0
\end{eqnarray}

\noindent
The expansions (\ref{v4}) must terminate at some power of $\rho$. If the expansions terminate at the first power of $\rho$, then all coefficients after $a_1$, $b_1$ being zero, the third pair of relations (\ref{v7}) give

\begin{equation}
\label{v8}
\frac{b_1}{a_1}=-\frac{\beta_1}{\beta}
\end{equation}

\noindent
We substitute this value of $b_1$ in terms of $a_1$ in the second pair of equations (\ref{v7}), we multiply the first of these by $\beta_1$ and the second by $\beta$ and subtract the one from the other. We obtain the relation

\begin{equation}
\label{v9}
\beta_1(1+i\alpha)+\beta\alpha+\beta_1(1+2k+i\alpha)-\frac{\beta_1}{\beta}\alpha=0
\end{equation}

\noindent
which gives the value of $k$

\begin{equation}
\label{v10}
k=-1-i\alpha+\frac{\alpha z}{\beta}
\end{equation}

\noindent
In general, if expansion (\ref{v4}) terminates in $\rho^n$ we have

\begin{equation}
\label{v11}
k=-n-i\alpha+\frac{\alpha z}{\beta}
\end{equation}

\noindent
we substitute (\ref{v8}) and (\ref{v10}) in the first pair of relations (\ref{v7}), we eliminate the coefficients $a_1$ and $b_1$, and obtain

\begin{equation}
\label{v12}
-\frac{\beta_2}{\beta}=\frac{\beta_1+3}{(\beta-3i)(-1-2i\alpha+\frac{2\alpha z}{\beta})}
\end{equation}

\noindent
which yields the complex equation of the second degree

\begin{equation}
\label{v13}
(2\alpha^2+6i\alpha)z^2+(-9-18i\alpha)z+9-20\alpha^2+12i\alpha=0
\end{equation}

The moduli of the two roots of this equation give the masses of two charged particles in units of the electron mass. These particles
have no nuclear interaction. The roots are

\begin{eqnarray}
\label{v14}
&&z_1=0.999892+i 5.18192\times10^{-7}\\ \nonumber
&&z_2=2.50009-i 205.546
\end{eqnarray}

\noindent
and their moduli are

\begin{eqnarray}
\label{v15}
&&|z_1|=0.999893\\ \nonumber
&&|z_2|=205.561=105.042 \textrm{MeV}
\end{eqnarray}

\noindent
The first root gives the mass of the electron in the field of a positive charge; the second root gives the mass of the muon in the field of a positive charge. The experimental value of the muon mass is 206.768 (105.659MeV).

Further values of masses of heavy leptons may be obtained by terminating expansions (\ref{v4}) at higher powers of $\rho$.

\section{The Nuclear Field and The Hadrons}

\subsection{The Potential $\Phi_3$}

Referring to table~\ref{tab1.1}, we see that the components of the field due to $\Phi_3$, are $F_{13}=-\frac{\partial \Phi_3}{\partial x_1}$ and $F_{23}=-\frac{\partial \Phi_3}{\partial x_2}$. The energy tensor (\ref{i3}) is

\begin{equation}
\label{vi1}
U^\nu_\mu=\frac{1}{2\pi}(-F^{\nu\alpha}F_{\mu\alpha}+\frac{1}{4}g^\nu_\mu F^{\alpha\beta}F_{\alpha\beta})
\end{equation}

\noindent
we calculate first $F^{\alpha\beta}F_{\alpha\beta}$. The only components of $F_{\mu\nu}$ being $F_{13}=-F_{31}$ and $F_{23}=-F_{32}$ , we have

\begin{eqnarray}
\label{vi2}
F^{\alpha\beta}F_{\alpha\beta}&&=2F^{13}F_{13}+2F^{23}F_{23}\\ \nonumber &&
=2[g^{11}g^{33}(F_{13})^2+g^{22}g^{33}(F_{23})^2]\\ \nonumber &&
=2g^{33}[g^{11}(F_{13})^2+g^{22}(F_{23})^2]\\ \nonumber &&
=2e^{-\gamma}r^{-2}\sin^{-2}\theta [e^{-\alpha}(F_{13})^2+e^{-\beta}r^{-2}(F_{23})^2]
\end{eqnarray}

\noindent
and

\begin{eqnarray}
\label{vi3}
&&F^{1\alpha}F_{2\alpha}=F^{13}F_{23}=g^{11}g^{33}F_{13}F_{23}=e^{-(\alpha+\gamma)}r^{-2}F_{13}F_{23}\\ \nonumber
&&F^{2\alpha}F_{1\alpha}=F^{23}F_{13}=g^{22}g^{33}F_{23}F_{13}=e^{-(\beta+\gamma)}r^{-2}F_{13}F_{23}\\ \nonumber
&&F^{13}F^{13}=g^{11}g^{33}(F_{13})^2=e^{-(\alpha+\gamma)}r^{-2}(F_{13})^2\\ \nonumber
&&F^{23}F^{23}=g^{22}g^{33}(F_{23})^2=e^{-(\beta+\gamma)}r^{-2}(F_{23})^2
\end{eqnarray}

\noindent
Also

\begin{eqnarray}
\label{vi4}
&&U^1_1=\frac{1}{4\pi}e^{-\gamma}r^{-2}\sin^{-2}\theta [+e^{-\alpha}(F_{13})^2+e^{-\beta}r^{-2}(F_{23})^2]\\ \nonumber
&&U^2_2=\frac{1}{4\pi}e^{-\gamma}r^{-2}\sin^{-2}\theta [-e^{-\alpha}(F_{13})^2-e^{-\beta}r^{-2}(F_{23})^2]\\ \nonumber
&&U^3_3=\frac{1}{4\pi}e^{-\gamma}r^{-2}\sin^{-2}\theta [-e^{-\alpha}(F_{13})^2-e^{-\beta}r^{-2}(F_{23})^2]\\ \nonumber
&&U^4_4=\frac{1}{4\pi}e^{-\gamma}r^{-2}\sin^{-2}\theta [+e^{-\alpha}(F_{13})^2+e^{-\beta}r^{-2}(F_{23})^2]\\ \nonumber
&&U^2_1=\frac{1}{2\pi}e^{-(\beta+\gamma)}r^{-4}\sin^{-2}\theta (F_{13}F_{23})\\ \nonumber
&&U^1_2=\frac{1}{2\pi}e^{-(\alpha+\gamma)}r^{-4}\sin^{-2}\theta (F_{13}F_{23})\\ \nonumber
&&U=U^1_1+U^2_2+U^3_3+U^4_4=0
\end{eqnarray}

\noindent
We remark that since $e^\alpha$, $e^\beta$, $e^\gamma$, $e^\delta$ are pure numbers, the expressions in square brackets show that $F_{13}$ and $\frac{1}{r}F_{23}$ have the same dimensions. This fact suggests an analogy with the field of a dipole, which is of the form

\begin{eqnarray}
\label{vi5}
H_r=\frac{2\mu \cos\theta}{r^3} \\ \nonumber
H_\theta=\frac{\mu \sin\theta}{r^2}
\end{eqnarray}

\noindent
deriving from a potential

\begin{equation}
\label{vi6}
\Phi=\frac{\mu \cos\theta}{r^2}
\end{equation}

We assume therefore the following expression for the potential $\Phi_3$

\begin{equation}
\label{vi7}
\Phi_3=\frac{\sigma \cos\theta}{(r+D)^2}
\end{equation}

\noindent
where $\sigma$ is the strength of the source, $D$ is a basic length introduced in analogy with formula (\ref{ii17}), and $\theta$ is the angle between the axis of the dipole and the radius vector to the point of observation. The field due to the potential $\Phi_3$ is

\begin{eqnarray}
\label{vi8}
F_{13}=-\frac{\partial \Phi_3}{\partial r}=\frac{-2\sigma \cos\theta}{(r+D)^3}\\
\label{vi9}
F_{23}=-\frac{\partial \Phi_3}{\partial \theta}=\frac{\sigma \sin\theta}{(r+D)^2}
\end{eqnarray}

The energy density, taken in covariant form, is

\begin{equation}
\label{vi10}
U_{33}=g_{33}U^3_3=-\frac{1}{4\pi}[-e^{-\alpha}(F_{13})^2-e^{-\beta}r^{-2}(F_{23})^2]
\end{equation}

\noindent
As a first approximation we assume Galilean coordinates, so that $U_{33}$ is equal to

\begin{equation}
\label{vi11}
U_{33}=\frac{1}{4\pi}[(F_{13})^2+r^{-2}(F_{23})^2]
\end{equation}

\noindent
Substituting (\ref{vi8}) and (\ref{vi9}) in (\ref{vi11}) we get

\begin{equation}
\label{vi12}
U_{33}=\frac{1}{4\pi}\sigma^2[\frac{4\cos^2\theta}{(r+D)^6}+\frac{\sin^2\theta}{r^2(r+D)^4}]
\end{equation}

\noindent
Integrating over all space we obtain the total energy

\begin{equation}
\label{vi13}
W=\frac{4}{15}\frac{\sigma^2}{D^3}
\end{equation}

To find the values of $\sigma$ and $D$, we first seek to find an order of magnitude for $\sigma$. We make the assumption that the field due to the potential $\Phi_3$ is a constituent of the nuclear field. We consider then the force between two protons in a nucleus. The field of a proton for $\theta=0$ being $2\sigma(r+D)^{-3}$, the energy of the second proton in the presence of the first is

\begin{equation}
\label{vi14}
V=-\frac{2\sigma^2}{(r+D)^3}
\end{equation}

\noindent
The nuclear interaction between the two protons will be

\begin{equation}
\label{vi15}
\frac{\partial V}{\partial r}=\frac{6\sigma^2}{(r+D)^4}
\end{equation}

The electric repulsion between the two protons is $e^2(r+A)^{-2}$, where $A$ is the basic length for the electric field $A=0.9393\times 10^{-13}\rm cm$, see (\ref{iv11}). We equate the two forces

\begin{equation}
\nonumber
\frac{6\sigma^2}{(r+D)^4}=\frac{e^2}{(r+A)^2}
\end{equation}

\noindent
If we adopt for $D$ the Compton wavelength for the proton $\lambdabar=0.21\times 10^{-13}\rm cm$, we obtain

\begin{equation}
\nonumber
\frac{\sigma}{e}=\frac{(r+0.21\times 10^{-13})^2\textrm{cm}}{\sqrt{6} (r+0.94\times 10^{-13})\textrm{cm}}
\end{equation}

\noindent
The distance $r$ between protons being of the order of a fermi, $\frac{\sigma}{e}$ will also be of the order of a fermi, and $\sigma$ will be of the order

\begin{equation}
\label{vi16}
\sigma\simeq 10^{-13} e\simeq 10^{-23}
\end{equation}

We can find the order of magnitude of the energy stored by the field in the whole space by inserting the values of $\sigma$ and $D$ in (\ref{vi13})

\begin{equation}
\label{vi17}
W\sim \frac{4\times 10^{-46}}{15\times(0.21\times 10^{-13})^3}\sim 10^{-5} \textrm{erg}
\end{equation}

The mass of proton being $M=1.6\times 10^{-13} \rm erg$, the ratio of the energy of the nuclear field to the mass of the proton will be of the order of $\frac{W}{M}\sim\frac{10^{-5}}{10^{-3}}\sim 10^{-2}\sim\alpha\simeq\frac{1}{137}$, that is, the fine structure constant. We equate therefore the expression (\ref{vi13}) to $\alpha Mc^2$

\begin{equation}
\nonumber
\frac{4}{15}\frac{\sigma^2}{\lambdabar^3}=\alpha Mc^2=\frac{e^2}{\hbar c} Mc^2=\frac{e^2}{\lambdabar}
\end{equation}

\noindent
so that $\sigma^2=3.75e^2\lambdabar^2$, or

\begin{equation}
\label{vi18}
\sigma=1.936e\lambdabar
\end{equation}

\noindent
For the value of $M$ which we shall use, we deduct from the mass $M_p=938.26\rm MeV$ of the proton the mass $m$ of its positive charge, which is of electromagnetic nature. We thus get

\begin{eqnarray}
\label{vi19}
&&M=938.26-0.51=937.75 \textrm{MeV} \\
\label{vi20}
&&\lambdabar=0.2104\times 10^{-13}\textrm{cm} \\
\label{vi21}
&& \sigma=1.9569\times 10^{-23}\textrm{cgs}
\end{eqnarray}

\subsection{The Baryons}

If we substitute the potential energy of the nuclear field (\ref{vi14}) in the wave equation (\ref{v1}), we must obtain the masses of baryons, following the same method that was used to obtain the leptons. The wave equations are

\begin{eqnarray}
\label{vii1}
&& (E-Mc^2-W)G+\hbar c \frac{dF}{dr}-\frac{\hbar c k}{r}F=0 \\ \nonumber
&& (E+Mc^2-W)F-\hbar c \frac{dG}{dr}-\frac{\hbar c k}{r}G=0
\end{eqnarray}

\noindent
We make the following substitutions in these equations

\begin{eqnarray}
\label{vii2}
&& z=\frac{E}{Mc^2} \\ \nonumber
&& \beta_1=1+z \\ \nonumber
&& \beta_2=1-z \\ \nonumber
&& \beta=\sqrt{\beta_1 \beta_2}=\sqrt{1-z^2} \\ \nonumber
&& \alpha=\frac{e^2}{\hbar c} \\ \nonumber
&& \rho=\frac{\beta r}{\lambdabar} \\ \nonumber
&& W=-\frac{2\sigma^2}{(r+\lambdabar)^3} \\ \nonumber
\end{eqnarray}

\noindent
where $W$ has been taken for $\theta=0$.

We take two new wave functions defined by

\begin{eqnarray}
\label{vii3}
f(\rho)=e^\rho F \\ \nonumber
g(\rho)=e^\rho G
\end{eqnarray}

\noindent
and obtain the two equations

\begin{eqnarray}
\label{vii4}
&& f^\prime-f-\frac{kf}{\rho}-(\frac{\beta_2}{\beta}-\frac{W}{\beta Mc^2})g=0 \\ \nonumber
&& g^\prime-g+\frac{kg}{\rho}-(\frac{\beta_1}{\beta}+\frac{W}{\beta Mc^2})f=0
\end{eqnarray}

\noindent
We transform the expression $\frac{W}{\beta Mc^2}$ using relations (\ref{vi14}) and (\ref{vi19})

\begin{eqnarray}
\nonumber
\frac{W}{\beta Mc^2}&&=-\frac{2\sigma^2}{(r+\lambdabar)^3}\frac{1}{\beta Mc^2} \\ \nonumber
&& =-\frac{7.5e^2\lambdabar^2}{(\frac{\lambdabar}{\beta})^3(\rho+\beta)^3\beta Mc^2} \\ \nonumber
&& =-\frac{7.5e^2 \beta^2}{\lambdabar (\rho+\beta)^3 Mc^2}
\end{eqnarray}

\noindent
Substituting $\frac{\hbar}{Mc}=\lambdabar$ and $\frac{e^2}{\hbar c}=\alpha$ we get

\begin{equation}
\label{vii5}
\frac{W}{\beta Mc^2}=-\frac{7.5\alpha\beta^2}{(\rho+\beta)^3}=-0.05473\frac{\beta^2}{(\rho+\beta)^3}
\end{equation}

\noindent
We denote the number $7.5\alpha$ by $p$.

\begin{equation}
\label{vii6}
p=7.5\alpha=0.05473
\end{equation}

\noindent
We take two new wave functions $\phi$ and $\psi$ defined by

\begin{eqnarray}
\label{vii7}
&& f=\rho^u \exp[\frac{s}{2(\rho+\beta)^2}] \phi \\ \nonumber
&& g=\rho^u \exp[\frac{s}{2(\rho+\beta)^2}] \psi
\end{eqnarray}

\noindent
Equations (\ref{vii4}) become

\begin{eqnarray}
\label{vii8}
&& \rho \phi^\prime+[u+\rho\frac{s}{(\rho+\beta)^3}-\rho-k]\phi-[\frac{\beta_2}{\beta}-\frac{p\beta^2}{(\rho+\beta)^3}]\rho\psi=0 \\ \nonumber
&& \rho \psi^\prime+[u+\rho\frac{s}{(\rho+\beta)^3}-\rho+k]\psi-[\frac{\beta_2}{\beta}+\frac{p\beta^2}{(\rho+\beta)^3}]\rho\phi=0
\end{eqnarray}

\noindent
We expand $\phi$ and $\psi$ in terms of $\rho$

\begin{eqnarray}
\label{vii9}
&& \phi=A_0+A_1\rho+A_2\rho^2+A_3\rho^3+\cdots \\ \nonumber
&& \psi=B_0+B_1\rho+B_2\rho^2+B_3\rho^3+\cdots
\end{eqnarray}

\noindent
Substituting these expansions in (\ref{vii8}), and equating to zero the constant terms, we get $u=k$, $B_0=0$, or $u=-k$, $A_0=0$. We choose

\begin{eqnarray}
\label{vii10}
&&u=-k \\ \nonumber
&&A_0=0
\end{eqnarray}

\noindent
Writing $\xi=\rho+\beta$ equations (\ref{vii8}) become

\begin{eqnarray}
\label{vii11}
&&(\xi-\beta)\phi^\prime+[-2k+\frac{(\xi-\beta)s}{\xi^3}-(\xi-\beta)]\phi-[\frac{\beta_2}{\beta}-\frac{p\beta^2}{\xi^3}](\xi-\beta)\psi=0 \\ \nonumber
&&(\xi-\beta)\psi^\prime+[\frac{(\xi-\beta)s}{\xi^3}-(\xi-\beta)]\psi-[\frac{\beta_1}{\beta}+\frac{p\beta^2}{\xi^3}](\xi-\beta)\phi=0
\end{eqnarray}

\noindent
We expand $\phi$ and $\psi$ in powers of $\xi$

\begin{eqnarray}
\label{vii12}
&& \phi=a_0+a_1\xi+a_2\xi^2+a_3\xi^3+\cdots \\ \nonumber
&& \psi=b_0+b_1\xi+b_2\xi^2+b_3\xi^3+\cdots
\end{eqnarray}

\noindent
We substitute these expressions in equations (\ref{vii11}) and equate constant terms to zero, and we obtain

\begin{eqnarray}
\nonumber
&&-sa_0-p\beta^2 b_0=0\\ \nonumber
&&-sb_0+p\beta^2 a_0=0
\end{eqnarray}

\noindent
which gives $s^2=-p^2\beta^4$, or $s=\pm ip\beta^2$.

We choose $s=ip\beta^2$ which gives

\begin{equation}
\label{vii13}
b_0=-i a_0
\end{equation}

\noindent
Equations (\ref{vii11}) become, after dividing the first by $(\xi-\beta)$ and multiplying both by $\xi^3$

\begin{eqnarray}
\label{vii14}
&&\xi^3(\xi-\beta)\phi^\prime+[-2k+ip\beta^2(\xi-\beta)-\xi^3(\xi-\beta)]\phi-[\frac{\beta_2}{\beta}\xi^3-p\beta^2](\xi-\beta)\psi=0\\ \nonumber
&&\xi^3\psi^\prime+[ip\beta^2-\xi^3]\psi-[\frac{\beta_1}{\beta}\xi^3+p\beta^2]\phi=0
\end{eqnarray}

\noindent
Equating the coefficients of $\xi$ to zero we get

\begin{equation}
\label{vii15}
b_1=-ia_1
\end{equation}

\noindent
Equating the coefficients of $\xi^2$ to zero we get

\begin{equation}
\label{vii16}
b_2=-ia_2
\end{equation}

We continue by equating the coefficients of $\xi^3$, $\xi^4$, $\xi^5$ and successive powers of $\xi$ to zero and obtain the following sequence of equations

\begin{eqnarray}
\label{vii17}
&& ip\beta^3 a_3+p\beta^3 b_3+\beta a_1+(2k-\beta+i\beta_2)a_0=0 \\ \nonumber
&& ip\beta^2 b_3-p\beta^2 a_3-i a_1+(i-\frac{\beta_1}{\beta})a_0=0 \\ \nonumber
&& ip\beta^3 a_4+p\beta^3 b_4-i p\beta^2 a_3-p\beta^2 b_3+2\beta a_2+(2k-1-\beta+i\beta_2)a_1+(1-i\frac{\beta_2}{\beta})a_0=0\\ \nonumber
&& ip\beta^2 b_4-p\beta^2 a_4-2 i a_2+(i-\frac{\beta_1}{\beta})a_1=0\\ \nonumber
&& ip\beta^3 a_5+p\beta^3 b_5-i p\beta^2 a_4-p\beta^2 b_4+3\beta a_3+(2k-2-\beta+i \beta_2)a_2+(1-i\frac{\beta_2}{\beta})a_1=0\\ \nonumber
&& ip\beta^2 b_5-p\beta^2 a_5+3 b_3+(i-\frac{\beta_1}{\beta})a_2=0\\ \nonumber
&& ip\beta^3 a_6+p\beta^3 b_6-i p\beta^2 a_5-p\beta^2 b_5+4\beta a_4+(2k-3-\beta)a_3+\beta_2 b_3+(1-i\frac{\beta_2}{\beta})a_2=0 \\ \nonumber
&& ip\beta^2 b_6-p\beta^2 a_6+4 b_4+b_3-\frac{\beta1}{\beta}a_3=0\\ \nonumber
&& ip\beta^3 a_7+p\beta^3 b_7-i p\beta^2 a_6-p\beta^2 b_6+5\beta a_5+(2k-4-\beta)a_4+\beta_2 b_4+\frac{\beta_2}{\beta}b_3+a_3=0\\ \nonumber
&& ip\beta^2 b_7-p\beta^2 a_7+5 b_5-b_4-\frac{\beta_1}{\beta}a_4=0\\ \nonumber
&& \cdot\cdot\cdot\cdot\cdot\cdot\cdot\cdot\\ \nonumber
&& \cdot\cdot\cdot\cdot\cdot\cdot\cdot\cdot\\ \nonumber
&& ip\beta^3 a_n+p\beta^3 b_n-ip\beta^2 a_{n-1}-p\beta^2 b_{n-1}+(n-2)\beta a_{n-2}+(2k-n+3-\beta)a_{n-3}\\ \nonumber
&& \quad\quad\quad\quad -\beta_2 b_{n-3}+\frac{\beta_2}{\beta}b_{n-4}+a_{n-4}=0 \\ \nonumber
&& ip\beta^2 b_n-p\beta^2 a_n+(n-2)b_{n-2}-b_{n-3}-\frac{\beta_1}{\beta}a_{n-3}=0
\end{eqnarray}

\subsubsection{Expansions (\ref{vii12}) terminate with $\xi^3$}

If the expansions (\ref{vii12}) terminate at $\xi^3$ so that $a_4$, $b_4$ and all coefficients after $a_4$ and $b_4$ vanish, the second equation of the fourth pair in (\ref{vii17}) will give

\begin{equation}
\label{vii18}
b_3=-\frac{\beta_1}{\beta}a_3
\end{equation}

Multiplying the first equation of the fourth pair in (\ref{vii17}) by $\frac{\beta_1}{\beta}$ and subtracting it from the second equation of the third pair in (\ref{vii17}), we find

\begin{equation}
\label{vii19}
\frac{\beta_1}{\beta}(6-2k)a_3=0
\end{equation}

\noindent
which has the two solutions

\begin{eqnarray}
\label{vii20}
\beta_1=0\\
k=3 \nonumber
\end{eqnarray}

The solution $\beta_1=0$, that is $z=-1$, represents a particle with a negative mass equal to the neutron mass. The second solution $k=3$ substituted in the first equation of the third pair in (\ref{vii17}) gives

\begin{equation}
\label{vii21}
-3\beta a_3+(-4+\beta-i \beta_2)a_2+(i\frac{\beta_2}{\beta}-1)a_1=0
\end{equation}

We eliminate $a_1$ between (\ref{vii21}) and the second equation of the second pair in (\ref{vii17}). Multiplying (\ref{vii21}) by $\frac{\beta_1}{\beta}$ and subtracting from second equation of the second pair in (\ref{vii17}) we obtain the equation

\begin{equation}
\label{vii22}
3\beta_1 a_3+(4\frac{\beta_1}{\beta}-\beta_1+i\beta-2i)a_2=0
\end{equation}

We have now to eliminate $a_2$ between (\ref{vii22}) and the second equation of the third pair in (\ref{vii17}). Multiplying this last equation by $\beta$ and taking account of relation (\ref{vii18}) and adding to (\ref{vii22}) we get

\begin{equation}
\label{vii23}
\beta(\frac{\beta_1}{\beta}-i)+4\frac{\beta_1}{\beta}-\beta_1+i\beta-2i=0
\end{equation}

\noindent
which may be written as

\begin{equation}
\label{vii24}
\beta(i+\beta_1)=\beta_1(2+i\beta_2)
\end{equation}

\noindent
Squaring, we get

\begin{equation}
\label{vii25}
(1-z^2)[-1+2i(1+z)+(1+z)^2]=(1+z)^2[4+2i(1-z)-(1-z)^2]
\end{equation}

\noindent
which reduces to the equation of the second order degree in $z$

\begin{equation}
\label{vii26}
(2-2i)z^2+3z+3+2i=0
\end{equation}

\noindent
Writing $z=x+iy$ and separating real and imaginary parts, we obtain the two simultaneous equations

\begin{eqnarray}
\label{vii27}
&&2x^2+4xy-2y^2+3x+3=0 \\ \nonumber
&&2x^2-4xy-2y^2-3y-2=0
\end{eqnarray}

\noindent
whose solutions are

\begin{eqnarray}
\label{vii28}
&&(x_1,y_1)=(0.23757,-1.16572)\\ \nonumber
&&(x_2,y_2)=(-0.98757,0.41572)
\end{eqnarray}

We consider only the first root whose real part is positive. Its modulus is $|z|=1.18968$. Multiplying this by the reduced mass of the nucleon $937.75 \rm MeV$ from (\ref{vi19}) we get

\begin{equation}
\nonumber
M=1115.62 \textrm{MeV}
\end{equation}

\noindent
which is to be compared with the mass of the baryon

\begin{equation}
\label{vii29}
\Lambda=1115.6\pm0.05 \textrm{MeV}
\end{equation}

\subsubsection{Expansions (\ref{vii12}) terminate with $\xi^4$}

In equations (\ref{vii17}), $a_5$, $b_5$ and all the following coefficients vanish. So the second equation of the fifth pair in (\ref{vii17}) gives

\begin{equation}
\label{vii30}
b_4=-\frac{\beta_1}{\beta}a_4
\end{equation}

Multiplying the second equation of the fourth pair in (\ref{vii17}) by $\frac{\beta_2}{\beta}$ and subtracting from the first equation of the fifth pair in (\ref{vii17}) we get

\begin{equation}
\label{vii31}
4-2k+4=0
\end{equation}

\noindent

which gives $k=4$.

Replacing $b_4$ from (\ref{vii30}) we get

\begin{equation}
\label{vii32}
a_4=-\frac{1}{4}(\frac{\beta}{\beta_1}b_3+a_3)
\end{equation}

Multiplying the second equation of the third pair in (\ref{vii17}) by $\frac{\beta_2}{\beta}$ and subtracting from the first equation of the fourth pair in (\ref{vii17}) we get

\begin{equation}
\label{vii33}
-4\beta a_4+(-5+\beta)a_3+(\beta_2-3\frac{\beta_2}{\beta})b_3=0
\end{equation}

\noindent
Replacing $a_4$ from (\ref{vii32}) into the second equation of the fourth pair in (\ref{vii17}) we get

\begin{equation}
\label{vii34}
b_3=-\frac{\beta}{\beta_2}\frac{5-2\beta}{3-2\beta}a_3
\end{equation}

\noindent
We replace in the second equation of the fourth pair in (\ref{vii17}) $b_3$ by its value from (\ref{vii34}) and we get

\begin{equation}
\label{vii35}
a_4=\frac{1}{2(3-2\beta)}a_3
\end{equation}

\noindent
We multiply the second equation of the second pair (\ref{vii17}) by $\frac{\beta_2}{\beta}$ and subtract from the first equation of the third pair in (\ref{vii17})

\begin{equation}
\label{vii36}
2p\beta(i\beta-z)a_4-3\beta a_3+(-6+\beta-i\beta_2+2i\frac{\beta_2}{\beta})a_2=0
\end{equation}

\noindent
We eliminate $a_2$ between (\ref{vii36}) and the second equation of the third pair in (\ref{vii6}) and get

\begin{equation}
\label{vii37}
2p\beta(i\beta-z)(i\beta-\beta_1)a_4-3(i\beta-\beta_1)a_3+3(i\beta_2-\beta+4)b_3=0
\end{equation}

\noindent
Replacing $a_4$ and $b_3$ by their values (\ref{vii35}) and (\ref{vii34}) we obtain the equation

\begin{equation}
\label{vii38}
\beta[p(-1+2z)+12i \beta_2+48]=\beta_2[ip(1+2z)+24i+12\beta_1]+60
\end{equation}

\noindent
Squaring and dividing by $12$, we obtain the equation of the third degree

\begin{eqnarray}
\label{vii39}
&&(4p+2ip+12-24i)z^3+(2p-12ip+24i)z^2\\ \nonumber
&&+(\frac{p^2}{12}-8p+3ip+36-96i)z-\frac{p^2}{12}+2p+7ip+102+96i=0
\end{eqnarray}

In a first approximation we neglect $p$ as found in (\ref{vii6}) $p=0.05473$. We have then after dividing by $6$, the equation to be solved

\begin{equation}
\label{vii40}
(2-4i)z^3+4iz^2+(6-16i)z+17+16i=0
\end{equation}

\noindent
We write $z=x+iy$ and obtain the two simultaneous equations

\begin{eqnarray}
\label{vii41}
&&2x^3-4y^3-12x^2y-3xy^2-8xy+6x+16y+17=0\\ \nonumber
&&2x^3+y^3-3x^2y-6xy^2-2x^2+2y^2-3y+8x-8=0
\end{eqnarray}

\noindent
It is possible to directly solve for the roots of (\ref{vii41}), however, numerical solutions to (\ref{vii39}) can be obtained exactly as

\begin{eqnarray}
\label{vii42}
&& (x_1,y_1)=(0.258,2.22)\\ \nonumber
&& (x_2,y_2)=(0.994,-1.15)\\ \nonumber
&& (x_3,y_3)=(-0.466,-1.46)
\end{eqnarray}

We consider only the first two solutions whose real parts are positive. The modulus of the first solution is $|z_1|=2.2332$ which when multiplied by the reduced mass of the nucleon $937.75\rm MeV$ as assessed in (\ref{vi19}), gives the mass

\begin{equation}
\nonumber
M_1=2094\textrm{MeV}
\end{equation}

\noindent
which is to be compared to the mass of the baryon

\begin{equation}
\label{vii43}
\Lambda=2100(-10,+20)\textrm{MeV}
\end{equation}

\noindent
The modulus of the second solution is $|z|=1.5225$ which gives the mass

\begin{equation}
\label{vii44}
M_2=1428\textrm{MeV}
\end{equation}

\noindent
to be compared with the mass of the baryon $N=1450\pm 32$.

\subsubsection{Expansions (\ref{vii12}) terminate with $\xi^5$}

Following the same procedure as above, and beginning with the second equation of the fifth pair in (\ref{vii17}), we find first $b_5=-\frac{\beta_1}{\beta}a_5$ and ending up with the second equation (\ref{vii11}), we arrive at the equation

\begin{eqnarray}
\label{vii45}
&&p\beta\{ [\beta(5-5z-10z^2-18i-25iz)+\beta_1(-13\beta_2+12z+5\beta_2 i+10\beta_2 iz)]\\ \nonumber
&&-2(\beta^2+i\beta z)a_4+2(i\beta^2-\beta z)\}+6\beta_1[\beta(i\beta_2+4)+\beta_2(-\beta_1-1)]a_4\\ \nonumber
&&+6[\beta(6\beta_2 i+\beta^2+28)+\beta_2(-9\beta_1-i\beta^2-7i)]b_4=0
\end{eqnarray}

In a first approximation, if we neglect $p$, we have the equation

\begin{eqnarray}
\label{vii46}
&&\beta_1[\beta(\beta_2 i+4)+\beta_2(-\beta_1-i)]a_4+\\ \nonumber
&&[\beta(6\beta_2 i+\beta^2+28)+\beta_2(-9\beta_1-i\beta^2-7i)]b_4=0
\end{eqnarray}

\noindent
Replacing $b_4$ in terms of $a_4$ derived from the second of the fourth equation and the fifth pair of equations in (\ref{vii17}), that is

\begin{equation}
\label{vii47}
b_4=-\frac{\beta}{\beta_1}\frac{3-\beta}{2-\beta}a_4
\end{equation}

\noindent
we arrive at the equation

\begin{eqnarray}
\nonumber
\beta&&[102-18z^2+i(29-29z-2z^2+2z^3)]=\\ \nonumber
&&\beta_2[65+65z-2z^2-2z^3+i(33-12z^2)]
\end{eqnarray}

\noindent
Squaring, we obtain the equation of the sixth degree

\begin{eqnarray}
\label{vii48}
(8-120i)z^6+36z^5&&+(84-1524i)z^4+648z^3+(978-1386i)z^2\\ \nonumber
&&+5931z+6427+1586i=0
\end{eqnarray}

The roots which have positive real parts are

\begin{eqnarray}
\nonumber
&&(x_1,y_1)=(1.502,-0.9444)\\ \nonumber
&&(x_2,y_2)=(0.1888,1.950)\\ \nonumber
&&(x_3,y_3)=(0.059,3.278)
\end{eqnarray}

The modulus of the first root is $|z_1|=1.77457$ giving, when multiplied by the mass of the nucleon $937.75$ (\ref{vi19})

\begin{equation}
\nonumber
M_1=1664\textrm{MeV}
\end{equation}

\noindent
to be compared with the mass of the baryon $\Delta=1650(-35,+45)\rm MeV$. The second root gives $|z_2|=1.95939$, that is

\begin{equation}
\nonumber
M_2=1837.4\textrm{MeV}
\end{equation}

\noindent
to be compared with the mass of the baryon $\Sigma=1840\pm 10 \rm MeV$. The third root gives $|z_3|=3.27879$

\begin{equation}
\nonumber
M_3=3074.7\textrm{MeV}
\end{equation}

\noindent
to be compared with the mass of the baryon $N=3030\textrm{MeV}$.

In similar fashion, by terminating the expressions (\ref{vii12}) at higher powers of $\xi$, we shall obtain equations of increasing degree in $z$. The roots of these equations will give successive values of the masses of baryons.

\subsection{The Mesons: The Klein-Gordon Equation}

We insert the potential energy $V=-\frac{\sigma^2}{(\rho+\lambdabar)^3}$ found in (\ref{vi14}), into the Klein-Gordon radial equation

\begin{equation}
\label{viii1}
[-\frac{1}{r^2}\frac{d}{dr}(r^2\frac{d}{dr})+\frac{l(l+1)}{r^2}]\phi=\frac{(E-V)^2-m^2 c^4}{\hbar^2 c^2}\phi
\end{equation}

\noindent
and make the change of variable $\rho=\frac{z}{\lambdabar}r$ where $z=\frac{E}{mc^2}$ and $\lambdabar=\frac{\hbar}{mc}$. We get the equation

\begin{equation}
\label{viii2}
\frac{z^2}{\lambdabar^2}[-\frac{1}{\rho^2}\frac{d}{d\rho}(\rho^2\frac{d}{d\rho})+\frac{l(l+1)}{\rho^2}]\phi=
\frac{(mc^2z+\frac{z^3}{\lambdabar^3}\frac{2\sigma^2}{(\rho+z)^3})^2 m^2 c^4}{\hbar^2 c^2}\phi
\end{equation}

\noindent
In (\ref{vi18}) we have found that $\sigma^2=3.75e^2\lambdabar^2$. Equation (\ref{viii2}) becomes therefore

\begin{equation}
\label{viii3}
\{z^2[\frac{d^2}{dz^2}+\frac{2}{\rho}\frac{d}{d\rho}-\frac{l(l+1)}{\rho^2}+(1+\frac{7.5\alpha z^2}{(\rho+z)^2})^2]-1\}\phi=0
\end{equation}

\noindent
where $\alpha=\frac{e^2}{\hbar c}$ is the fine-structure constant.

We write $\phi=e^{-\frac{1}{2}\rho}F$ and get

\begin{equation}
\label{viii4}
z^2\{F^{\prime\prime}+(\frac{2}{\rho}-1)F^\prime+[\frac{1}{4}-\frac{1}{\rho}-\frac{l(l+1)}{\rho^2}+(1+\frac{7.5\alpha z^2}{(\rho+z)^3})^2]F\}-F=0
\end{equation}

\noindent
We write next $F=\rho^s f$ and get

\begin{eqnarray}
\label{viii5}
&&z^2\{\rho^2 F^{\prime\prime}+[2(s+1)\rho-\rho^2]f^\prime+[s(s+1)-(s+1)\rho+\frac{\rho^2}{4}-l(l+1)\\ \nonumber
&&\quad\quad\quad\quad +\rho^2(1+\frac{7.5 \alpha z^2}{(\rho+z)^3})^2]f\}-\rho^2 f=0
\end{eqnarray}

\noindent
Equating constant terms to zero we have

\begin{equation}
\label{viii6}
s(s+1)-l(l+1)=0
\end{equation}

\noindent
which gives $s=l$ and $s=-l-1$. We here take $s=l$ for regular solutions.

Equation (\ref{viii5}) becomes therefore after dividing by $\rho$

\begin{eqnarray}
\label{viii7}
&& z^2\{\rho f^{\prime\prime}+[2(l+1)-\rho]f^\prime+[-(l+1)+\frac{\rho}{4} \\ \nonumber
&& \quad\quad\quad\quad +\rho(1+\frac{7.5\alpha z^2}{(\rho+z)^3})^2]f\}-\rho f=0
\end{eqnarray}

\noindent
We multiply throughout by $(\rho+z)^6$

\begin{eqnarray}
\label{viii8}
&& z^2(\rho+z)^6\{\rho f^{\prime\prime}+[2(l+1)-\rho]f^\prime+[-(l+1)+\frac{5}{4}\rho]f\} \\ \nonumber
&& \quad\quad\quad\quad +[15\alpha z^4(\rho+z)^3\rho+56.25\alpha^2 z^6 \rho-(\rho+z)^6\rho]f=0
\end{eqnarray}

\noindent
We expand $f$ in powers of $\rho$

\begin{equation}
\label{viii9}
f=a_0+a_1\rho+a_2\rho^2+a_3\rho^3+\cdots
\end{equation}

\noindent
We equate constant terms in equation (\ref{viii8}) to zero

\begin{equation}
\label{viii10}
z^8[2(l+1)a_1-(l+1)a_0]=0
\end{equation}

\noindent
which gives $a_1=\frac{1}{2}a_0$.

We equate to zero the coefficients of $\rho$ in equation (\ref{viii8})

\begin{equation}
\label{viii11}
z^8[2(2l+3)a_2-(l+2)a_1+\frac{5}{4}a_0]+(15\alpha z^7+56.25 \alpha^2 z^6-z^6)a_0=0
\end{equation}

\noindent
Taking the account of (\ref{viii10}) this equation becomes

\begin{equation}
\label{viii12}
z^2\{2(2l+3)a_2 z^2+[-(l-\frac{1}{2})z^2+30\alpha z+2(56.25\alpha^2-1)]a_1\}=0
\end{equation}

\noindent
which gives

\begin{equation}
\label{viii13}
a_2=\frac{1}{2(2l+3)}[l-\frac{1}{2}-\frac{30\alpha}{z}+\frac{2(1-56.25\alpha^2)}{z^2}]a_1
\end{equation}

If expansion (\ref{viii9}) terminates so that $a_2$ and following coefficients vanish, we have the equation

\begin{equation}
\label{viii14}
(l-\frac{1}{2})z^2-30\alpha z+2(1-56.25 \alpha^2)=0
\end{equation}

\noindent
that is

\begin{equation}
\label{viii15}
(l-\frac{1}{2})z^2-0.21892 z+1.994=0
\end{equation}

For $l=0$ from (\ref{viii14}) we have the equation of the second degree

\begin{equation}
\label{viii16}
z^2+0.43784z-3.988=0
\end{equation}

\noindent
The positive root is $z=1.79004$. This value, multiplied by the reduced mass of the nucleon, $937.75\rm MeV$ as assessed in (\ref{vi19}) gives the mass

\begin{equation}
\label{viii17}
M=1679 \textrm{MeV}
\end{equation}

\noindent
to be compared with the meson $g=1680\pm 20 \rm MeV$.

We equate next to zero the coefficient of $\rho^2$ in (\ref{viii8})

\begin{eqnarray}
\label{viii18}
&&6(l+2)a_3z^8+[-(l+3)a_2+\frac{5}{4}a_1]z^3 \\ \nonumber
&&\quad\quad +[12(2l+3)a_2-6(l-\frac{1}{2})a_1+15\alpha a_1]z^7 \\ \nonumber
&&\quad\quad +[90\alpha+56.25\alpha^2-1]a_1 z^6-12a_1 z^5=0
\end{eqnarray}

\noindent
We substitute the expression for $a_2$ found in (\ref{viii13})

\begin{equation}
\label{viii19}
a_2=\frac{1}{2(2l+3)}[l-\frac{1}{2}-\frac{30\alpha}{z}+\frac{2(1-56.25\alpha^2)}{z^2}]a_1
\end{equation}

\noindent
Dividing by $z^5$ we find

\begin{eqnarray}
\label{viii20}
&& 6(l+2)a_3 z^3+\{[-\frac{(l+3)(l+\frac{1}{2})}{2(2l+3)}+\frac{5}{4}]z^3 \\ \nonumber
&& \quad\quad +\frac{l+2}{2l+3}45\alpha z^2-[\frac{3(l+2)}{2l+3}(1-56.25\alpha)^2-90\alpha]z-675\alpha^2\}a_1=0
\end{eqnarray}

If in expansion (\ref{viii9}), $a_3$ and following coefficients vanish, we have the equation

\begin{eqnarray}
\label{viii21}
&&\frac{(2-l)(2l+9)}{4(2l+3)}z^3+0.32838\frac{l+2}{2l+3}z^2 \\ \nonumber
&&\quad\quad\quad\quad -(2.991\frac{l+2}{2l+3}+0.65676)z-0.036=0
\end{eqnarray}

For $l=0$, equation (\ref{viii21}) becomes

\begin{equation}
\label{viii22}
z^3+0.14595z^2-1.76714z-0.024=0
\end{equation}

\noindent
which has the positive root $z=1.2641$, which multiplied by the reduced mass of the nucleon $937.75\rm MeV$, gives the mass

\begin{equation}
\label{viii23}
M=1185\textrm{MeV}
\end{equation}

\noindent
to be compared with the meson $\epsilon=1200\pm 100 \rm MeV$.

We equate next to zero the coefficients of $\rho^3$ in (\ref{viii8})

\begin{eqnarray}
\label{viii24}
&&4(2l+5)z^4 a_4+[-(l+4)a_3+\frac{5}{4}a_2]z^4 \\ \nonumber
&&\quad\quad +[36(l+2)a_3+(15\alpha-6(l+3))a_2+\frac{15}{2}a_1]z^3 \\ \nonumber
&&\quad\quad +[(30(2l+3)56.25\alpha^2-1)a_2+(45\alpha-15(l+2)+37.5)a_1]z^2 \\ \nonumber
&&\quad\quad +(90\alpha-6)a_1 z-30a_1=0
\end{eqnarray}

\noindent
We substitute the expressions for $a_2$ and $a_3$ from (\ref{viii13}) and (\ref{viii20}), and obtain the relation

\begin{equation}
\label{viii25}
50.52631z^4a_4+[z^4+0.04605z^3-0.70615z^2-16.48146z-1.06517]a_1=0
\end{equation}

If $a_4$ and following coefficients in expansion (\ref{viii9}) vanish, relation (\ref{viii25}) gives an equation whose positive root is $z=2.641$, which multiplied by the reduced mass of the nucleon as assessed in (\ref{vi19}), gives the mass

\begin{equation}
\nonumber
M=2476\textrm{MeV}
\end{equation}

\noindent
to be compared with the meson $X=2500\pm32 \rm MeV$.

Further values of meson masses may be obtained by terminating expansion (\ref{viii9}) at higher powers of $\rho$.

\section{Conclusions}

In this monograph, the premise is set forth, that the difficulties which to day beset particle physics, are due to our adherence to the concept of point-like particles. A departure from that concept is proposed, namely, that we base our quest on the postulate that particles are infinitely extended in space.

An interpretation of general relativity that differs in some respects from that generally accepted, results in the formulation of expressions for the gravitational, electric, and nuclear potentials, which include each a basic length, and which are free from the infinities which occur in the integrations for the self energy. The particle is seen to be spread over all space, and its potentials are not interpreted in terms of a curvature of space, but as local stresses in the unit mesh at a point.

When the potentials are inserted in the wave equations they yield in a simple manner the masses of the muon, of baryons, and of mesons, as summarized in Table III.

\begin{table}
\caption{List of Particles (mass in MeV).}
\begin{ruledtabular}
\begin{tabular}{lll}
          & Calculated & Compared With \\
\hline
Leptons   & 105.04     & Muon 105.66   \\
\hline
          & 1115.62    & $\Lambda\quad 1115.6\pm 0.05$ \\
          & 2094       & $\Lambda\quad 2100(-10,+20)$\\
Baryons   & 1428       & $N\quad 1450\pm 32$ \\
          & 1664       & $\Delta\quad 1650(-35,+45)$\\
          & 1837       & $\Sigma\quad 1840\pm 10$ \\
          & 3074       & $N\quad 3030$\\
\hline
          & 1679       & $g\quad 1680\pm20$ \\
Mesons    & 1185       & $\epsilon\quad 1200\pm100$\\
          & 2476       & $X\quad 2500\pm32$
\end{tabular}
\end{ruledtabular}
\end{table}

\end{document}